\documentclass[runningheads]{llncs}
\usepackage{graphicx}
\usepackage{mathrsfs}
\usepackage{amsfonts,amssymb}
\usepackage{bm}
\usepackage{marvosym}

\usepackage{url}

\renewcommand\UrlFont{\color{blue}\rmfamily}

\begin{document}
\title{Model-based Convolutional De-Aliasing Network Learning for Parallel MR Imaging}
\titlerunning{Model-based Convolutional De-Aliasing Network Learning for pMRI}

%
\renewcommand\UrlFont{\color{blue}\rmfamily}

	%

\author{Yanxia Chen\inst{*1,2} \and
	Taohui Xiao\inst{1} \and
	Cheng Li\inst{1} \and
	Qiegen Liu\inst{3} \and
	Shanshan Wang\inst{1}\textsuperscript{(\Letter)}}
%

%

%
%
\institute{Paul C. Lauterbur Research Center for Biomedical Imaging,Shenzhen Institutes of Advanced Technology, Chinese Academy of Sciences, Shenzhen,Guangdong, China\\
	\email{sophiasswang@hotmail.com}\\ \and
	University of Chinese Academy of Sciences, Beijing, China \and
	Department of Electronic Information Engineering,Nanchang University, \\
	Nanchang, China}

\authorrunning{Y. Chen et al.}
%
%
\maketitle              
\begin{abstract}
Parallel imaging has been an essential technique to accelerate MR imaging. Nevertheless, the acceleration rate is still limited due to the ill-condition and challenges associated with the undersampled reconstruction. In this paper, we propose a model-based convolutional de-aliasing network with adaptive parameter learning to achieve accurate reconstruction from multi-coil undersampled k-space data. Three main contributions have been made: a de-aliasing reconstruction model was proposed to accelerate parallel MR imaging with deep learning exploring both spatial redundancy and multi-coil correlations; a split Bregman iteration algorithm was developed to solve the model efficiently; and unlike most existing parallel imaging methods which rely on the accuracy of the estimated multi-coil sensitivity, the proposed method can perform parallel reconstruction from undersampled data without explicit sensitivity calculation. Evaluations were conducted on \emph{in vivo} brain dataset with a variety of undersampling patterns and different acceleration factors. Our results demonstrated that this method could achieve superior performance in both quantitative and qualitative analysis, compared to three state-of-the-art methods.
\keywords{Parallel MR imaging \and Deep Learning \and Bregman iteration.}
\end{abstract}
\section{Introduction}
Magnetic Resonance Imaging (MRI) is an important imaging modality for both research and clinical uses. However, its slow imaging speed has limited its wide applications. Several approaches have been tried to accelerate MR scans \cite{ref_article1,ref_article2,ref_article3,ref_article4}, including fast sequence , parallel Magnetic Resonance Imaging (pMRI), and MR image reconstruction from undersampled k-space data. Of these approaches, pMRI simultaneously samples k-space data through a multichannel RF receiver coil array and combines the sensitivity information of the coil with gradient coding to reduce the number of samples required for reconstruction. 

Traditional pMRI reconstruction algorithms have achieved reliable reconstruction results using coil sensitivity information. Examples include sensitivity encoding (SENSE)~\cite{ref_article1}, and generalized autocalibrating partially parallel acquisitions (GRAPPA)~\cite{ref_article2}. In these algorithms, a small sensitivity error may introduce visible artifacts in the reconstructed image~\cite{ref_article3}. In addition to the challenge associated with coil sensitivity estimation, reconstruction from undersampled pMRI data is an ill-posed inverse problem. To address this issue, prior knowledge is often required. Typical methods include simultaneous autocalibrating and k-space estimation (SAKE)~\cite{ref_article5}, wavelet and total variation (TV) filtering~\cite{ref_article6,ref_article7}, and calibration-free and joint-sparse codes (LINDBERG)~\cite{ref_article8}.These methods have made encouraging progresses in fast MR imaging, Nevertheless, the optimization tends to be time consuming and its parameters are hard to adjust~\cite{ref_article9}.

In recent years, deep learning has shown encouraging capability to accelerate MR scan. Specifically, deep learning was first integrated with fast MR imaging in~\cite{ref_ISBI10} where a Convolutional Neural Network (CNN) model was trained to learn an end-to-end mapping between the undersampled zero-filled images and fully-sampled k-space data. Subsequently, a number of methods have been developed. These include deep cascade convolutional neural networks (DC-CNNs) in~\cite{ref_article11} for dynamic imaging. A variational network (VN)-based reconstruction approach was proposed in~\cite{ref_article9} to achieve fast, high-quality reconstruction from undersampled k-space data. ADMM-Net in~\cite{ref_article12} took the sampled k-space data as input and obtained accurate reconstruction results. Multi-channel generative adversarial network was used for parallel MRI reconstruction in K-space~\cite{ref_MICCAI13}. Model-Based Deep Learning (MoDL) was proposed in~\cite{ref_article14}  for image reconstruction using a deep learned prior, which combines the power of data-driven learning with the physics derived model-based framework.  MoDL has obtained promising performances. Nevertheless, this method also has some limitations that 1) the multi-coil sensitivity needs to be explicitly calculated and 2) the redundancy and correlations of the multi-coil data have not been thoroughly investigated.

Based on the above observation, this paper proposes a reconstruction model to further explore the redundant and correlations with deep learning, where split Bregman is used to develop an end-to-end learning and reconstruction algorithm that can avoid explicit calculation of the coil sensitivities. Specifically, the main contributions are summarized as follows: (1) a de-aliasing reconstruction model was proposed to accelerate parallel MR imaging with deep learning exploring both spatial redundancy and multi-coil correlations; (2) a split Bregman iteration algorithm was developed to solve the model efficiently; (3) and unlike most existing parallel imaging methods which rely on the accuracy of the estimated multi-coil sensitivity, the proposed method can perform parallel reconstruction from undersampled data without explicit sensitivity calculation.  With the proposed method, faster reconstruction speed and more stable reconstruction performance were achieved compared to classical iterative self-consistent parallel imaging reconstruction (SPIRiT)~\cite{ref_article3}, SAKE~\cite{ref_article5}, and the state-of-the-art MoDL~\cite{ref_article14} methods using \emph{in vivo} dataset. The source code for the proposed method will be made public available based on its acceptance.
 \begin{figure}
	\centering
	\includegraphics[width=90mm, height=45mm]{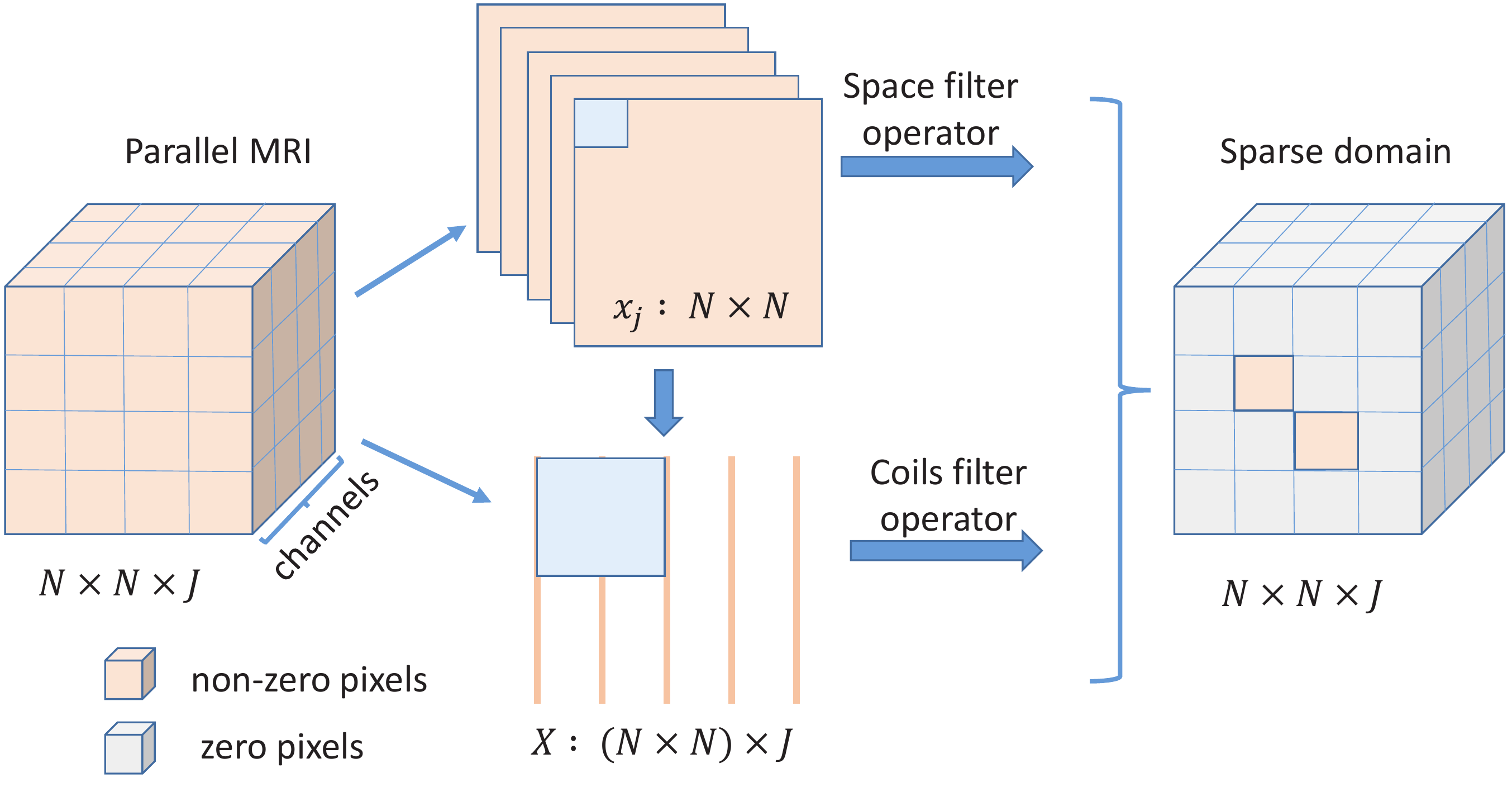}
	\caption{An illustration of the filter operator with convolutional neural networks for both spatial and multi-coil correlations.} \label{fig1}
\end{figure}
\section{Method}
Let $\textbf{A=MF}\in\mathbb{C}^{M \times N}$ denote a measurement matrix, where $\textbf{M}$ is an undersampled matrix and $\textbf{F}$ is the 2D Fourier transform. Our purpose is to reconstruct $\textbf{X}$ from the undersampled k-space data $\textbf{Y}\in\mathbb{C}^{M \times J}(M<<N)$, $\textbf{X}=\{\bm{x_1|x_2|\dots x_j|,\dots x_J|}\}\in\mathbb{C}^{N \times J}$, $|$ stacks the vectors as columns and J is the total number of receiver coils. This problem can be described by the following unconstrained problem:
\begin{equation}
\mathop{\arg\min}_{\textbf{X}}\left\{\frac{1}{2} \|\textbf{AX} - \textbf{Y}\|_2^2+\lambda \mathscr{P}(\mathbf{\Phi}{\textbf{X}}) \right\} \label{Eq1}
\end{equation}
where $\lambda$ is the regularization parameter, $\mathscr{P}(\cdot)$ is usually a regularization function derived from data priors, e.g., sparse prior and $\mathbf{\Phi}$ is learning filtering operator.
%

In general, the filter operator $\mathbf{\Phi}$ in Eq. (1) is fixed, e.g., Discrete Cosine Transform (DCT), TV and Discrete Wavelet Transform (DWT). In this work, we used CNN to adaptively learn the filter operator to obtain the maximum sparsity in the transform domain, in order to reduce the number of samples collected as much as possible. Eq. (1) can be rewritten as:
\begin{equation}
\mathop{\arg\min}_{\textbf{X}}\left\{\frac{1}{2} \|\textbf{AX} - \textbf{Y}\|_2^2+\lambda_s\mathscr{P}(\mathbf{\Phi_s}{\textbf{X}})+\lambda_{coils}\mathscr{P}(\mathbf{\Phi_{coils}}{\textbf{X}}) \right\} \label{Eq2}
\end{equation}
where $\frac{1}{2} \|\textbf{AX} - \textbf{Y}\|_2^2$ is the data fidelity term,  $\lambda_s\mathscr{P}(\mathbf{\Phi_s}{\textbf{X}})$ and $\lambda_{coils}\mathscr{P}(\mathbf{\Phi_{coils}}{\textbf{X}})$. are the regularization terms with two weighting parameters. $\mathbf{\Phi_s}$ represents the filter operator in the spatial domain and $\mathbf{\Phi_{coils}}$ represents the filter operator on the multi-coil. The illustration of the filter operator in Eq. (2) is shown in Fig.~\ref{fig1}.

\begin{figure}
	\includegraphics[width=\textwidth]{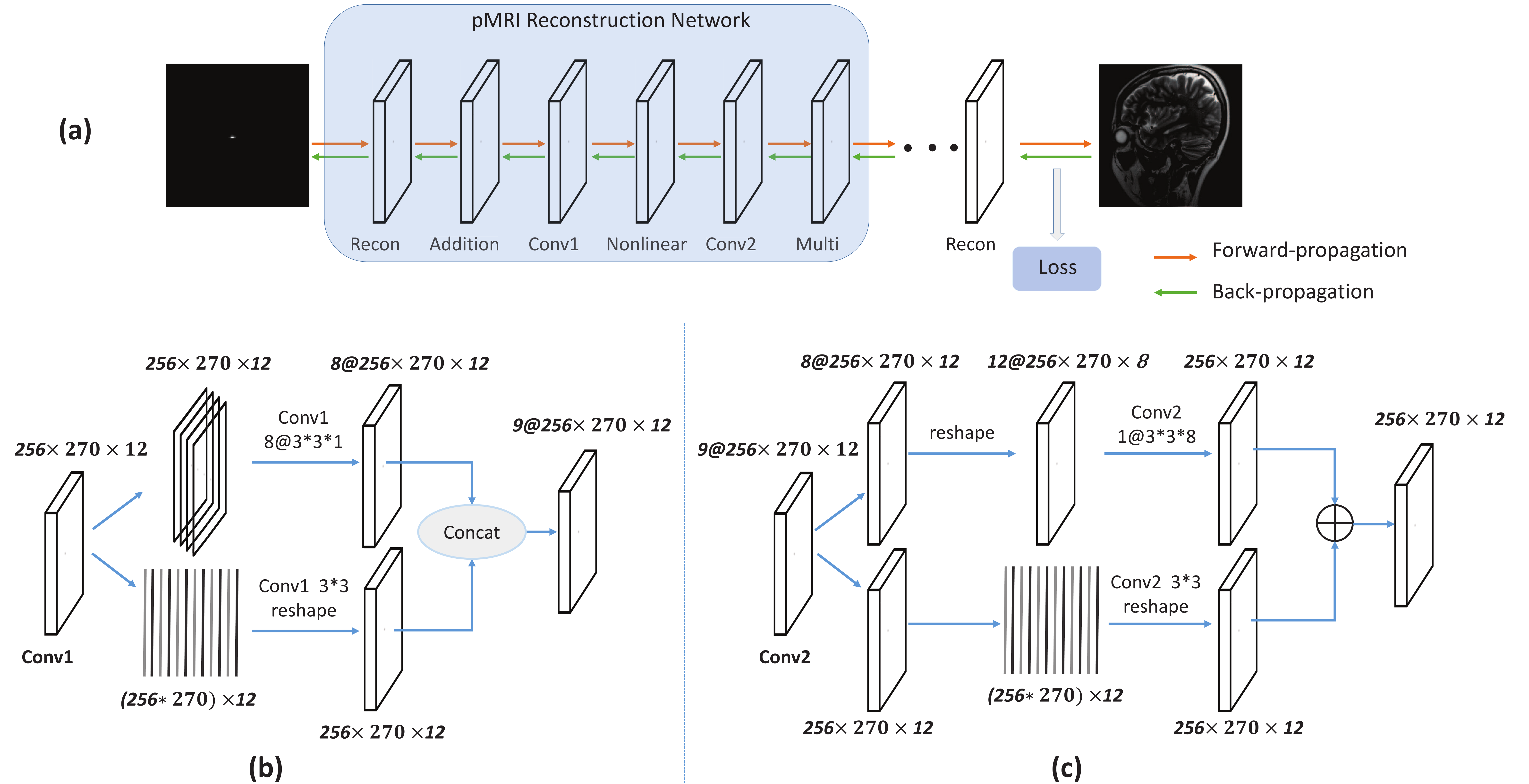}
	\caption{The proposed convolutional de-aliasing network architecture for pMRI reconstruction. (a) is the flow chart. The orange arrow indicates the process of reconstructing the undersampled k-space data by forward propagation, and the green arrow indicates the parameter updating through back propagation. (b) and (c) are the detailed configurations of $Conv1$ and $Conv2$.} \label{fig2}
\end{figure}

For convenience, we combine the two regularizations of the above formula.
\begin{equation}
\mathop{\arg\min}_{\textbf{X}}\left\{\frac{1}{2} \|\textbf{AX} - \textbf{Y}\|_2^2+\sum_{l=1}^L\lambda_l\mathscr{P}(\mathbf{\Phi}_l{\textbf{X}}) \right\} \label{Eq3}
\end{equation}
where $L$ represents the number of filters. Introducing auxiliary variables $\textbf{V}=\{\bm{v_1,v_2,\dots v_j,\dots v_J}\}$, Eq. (3) is equivalent to:  
\begin{equation}
\mathop{\arg\min}_{\textbf{X,V}}\left\{\frac{1}{2} \|\textbf{AX} - \textbf{Y}\|_2^2+\sum_{l=1}^L\lambda_l\mathscr{P}(\bm{\Phi_l} \textbf{V}) \right\}\ \ \ \ \ \ \ \ s.t.\ \textbf{X}=\textbf{V}
\end{equation}
Eq. (4) can be further changed into the unconstrained optimization formulation
\begin{equation}
\mathop{\arg\min}_{\textbf{X,V}}\left\{\frac{1}{2} \|\textbf{AX} - \textbf{Y}\|_2^2+\sum_{l=1}^L\lambda_l\mathscr{P}(\mathbf{\Phi}_l \textbf{V})+\frac{\rho}{2}\|\textbf{X}-\textbf{V}\|_2^2\right\}
\end{equation}
where $\rho$ denotes the penalty parameter. Let $\mathscr{K}=\frac{1}{2} \|\textbf{AX} - \textbf{Y}\|_2^2+\sum_{l=1}^L\lambda_l\mathscr{P}(\mathbf{\Phi}_l \textbf{V})$. Applying the split Bregman iteration~\cite{ref_article15}, Eq. (5) can be further solved with the following three simple iterations.
\begin{equation}
\left\{ \begin{array}{lll}
\textbf{X}^{k+1}=\mathop{\min}\limits_{\textbf{X}} \frac{1}{2}\|\textbf{AX} - \textbf{Y}\|_2^2+\frac{\rho}{2}\|\textbf{X}-\textbf{V}^k-b^k\|_2^2\\
\textbf{V}^{k+1}=\mathop{\min}\limits_{\textbf{V}}\sum_{l=1}^L\lambda_l\mathscr{P}(\mathbf{\Phi}_l \textbf{V})+\frac{\rho}{2}\|\textbf{X}^{k+1}-\textbf{V}-b^k\|_2^2\\
b^{k+1}=b^{k}+\textbf{V}^{k+1}-\textbf{X}^{k+1}
\end{array}
\right.
\end{equation}
where $b^k$ is an auxiliary variable. Let $\mu_1=(1-\alpha_r\rho)$, $\mu_2=\alpha_r\rho$ and $\tilde\lambda_l=\alpha_r\lambda_l$, $\alpha_r$ is the step size. We solve $\textbf{X}^{k+1}$ and $\textbf{V}^{k+1}$ in Eq. (6) using the least squares method and the gradient descent method, respectively. Then we have the following solutions:
\begin{equation}
\left\{ \begin{array}{lll}
\textbf{X}^{(n)}=\textbf{F}^T(\textbf{M}^T\textbf{M}+\rho^{(n)}\textbf{I})^{-1}[\textbf{M}^T\textbf{Y}+\rho^{(n)}\textbf{F}(\textbf{V}^{n-1}-b^{n-1})] \\
\textbf{V}^{(n,k)}=\mu_1\textbf{V}^{(n,k-1)}+\mu_2(\textbf{X}^{(n)}+b^{(n-1)})-\sum_{l=1}^L\tilde\lambda_l\mathbf{\Phi}_l^T\mathscr{F}(\mathbf{\Phi}_l \textbf{V}^{(n,k-1)})\\
b^{k+1}=b^k+\tilde\eta(\textbf{X}^{(n)}-\textbf{V}^{(n)}) \ \ \ \ \ \  {\forall}\ k\geq1.
\end{array}
\right.
\end{equation}
where $\mathscr{F}$ is the gradient of regularization function $\mathscr{P}(\cdot)$, and the parameter $\tilde\eta$ is an update rate. $n$ represents the $n-th$ iteration. $(n,k)$ represents the repetition of the substage $k$ times in the nth iteration. Eqn. (7) can be further written as:
\begin{equation}
\left\{ \begin{array}{lll}
Recon:\ \textbf{X}^{(n)}=\textbf{F}^T(\textbf{M}^T\textbf{M}+\rho^{(n)}\textbf{I})^{-1}[\textbf{M}^T\textbf{Y}+\rho^{(n)}\textbf{F}(\textbf{V}^{(n-1)}-b^{(n-1)})] \\
Addition:\ \textbf{V}^{(n,k)}=\mu_1\textbf{V}^{(n,k-1)}+\mu_2(\textbf{X}^{(n)}+b^{(n-1)})-\textbf{C}_2^{(n,k)}\\
Conv1:\ \textbf{C}_1^{(n,k)}=\sum_{l=1}^L(\bm{w_{1,l}^{(n,k)}}\times \textbf{V}^{(n,k-1)}+\bm{b_{1,l}^{(n,k)}})\\
Nonlinear:\ \bm{h^{(n,k)}}=\bm{S}_{PLF}(\textbf{C}_1^{(n,k)};\left\{\bm{p}_i,\bm{q}_i^{(n,k)}\right\}_{i=1}^{N_c})\\
Conv2:\textbf{C}_2^{(n,k)}=\sum_{l=1}^L(\bm{w_{2,l}^{(n,k)}}\times \bm{h^{(n,k)}}+\bm{b_{2,l}^{(n,k)}})\\
Multi:\ b^{(n)}=b^{(n-1)}+\tilde\eta(\textbf{X}^{(n)}-\textbf{V}^{(n)})\\
\end{array}
\right.
\end{equation}
 $\sum_{l=1}^L\tilde\lambda_l\mathbf{\Phi}_l^T\mathscr{F}(\mathbf{\Phi}_l \textbf{V}^{(n,k-1)})$ in Eq. (7) can be regarded as $f_{CNN2}\times \mathscr{F}(f_{CNN1}\times \textbf{V}^{(n,k-1)})$, where $f_{CNN1}$ represents the $Conv1$ layer for feature extraction. $\mathscr{F}(\cdot)$ represents the $nonlinear$ layer, approximated by piecewise linear functions. $f_{CNN2}$ represents the $Conv2$ layer for feature fusion. $\bm{S}_{PLF}(\cdot)$ is a piecewise linear function and $\left\{\bm{p}_i,\bm{q}_i^{(n,k)}\right\}_{i=1}^{N_c}$ is a set of control points.
%
%

The flow chart and the main architecture of the network are shown in Fig.~\ref{fig2} (a), where each layer corresponds to the above Eq. (8) and $Recon$ stands for reconstruction layer $\textbf{X}^{(n)}$. $Addition$ denotes the addition layer $\textbf{V}^{(n,k-1)}$. $Conv1$ and $Conv2$ denote the convolution layers $\textbf{C}_1^{(n,k)}$ and $\textbf{C}_2^{(n,k)}$ (see Fig.~\ref{fig2} (b) and (c)). $Nonlinear$ denotes the nonlinear layer $\bm{h^{(n,k)}}$. $Multi$ denotes the multiplier update layer $b^{(n)}$. The loss function is calculated between the reconstructed pMRI and the original fully sampled pMRI by a standard mean squared error.
\section{Experiments}
\subsubsection{Dataset and Experimental Setting}
Our method was evaluated using a 2D multichannel MR brain dataset collected from a 3T scanner (SIEMENS MANGETOM Trio Tim) with a 12-channel head coil and a \emph{Turbo-Spin-Echo (TSE)} sequence. The scanning parameters were TR = 2500ms, TE = 149ms and voxel resolution = $ 0.9\times 0.9\times 0.9 $mm. We randomly selected 80 images for training, 20 images for validation and 50 images for testing. Training and testing were implemented on an Ubuntu 16.04 LTS (64-bit) operating system equipped online model training took 45 hours on an Intel Xeon (R) CPU E5-2640 V4 @ 2.40GHz$ \times 40$, 64G. 

Four different undersampling patterns were used to evaluate the effectiveness of the proposed method at different acceleration factors. For forward propagation, we initialized the spatial domain $\mathbf{\Phi_s}$ with DCT filter operator of size $ 3 \times 3 \times 8$ and multi-coil $\mathbf{\Phi_{coils}}$ with TV filter operator of size $ 3 \times 3 \times 1$. Other parameters in the network were initialized as follows: filter numbers $(L=9)$, penalty parameter $(\rho=0.2)$, step size $(\alpha_r=0.3)$, $\mu_2=\rho \times \alpha_r=0.06$, $\mu_1=1-\mu_2=0.94$, $\tilde\eta=1.8$, number of iterations $(n=13)$, substage size $(k=1)$, batchsize $=1$, learning rate $=0.01$ and epoch $=400$. In the back propagation process, we updated all of the above parameters, including the penalty parameter $\rho$ in the $Recon$ layer, parameters $\mu_1$ and $\mu_2$ in the $Addition$ layer, filters $\bm{w_1}$ and bias $\bm{b_1}$ in the $Conv1$ layer, parameter $\left\{\bm{p}_i,\bm{q}_i^{(n,k)}\right\}$ in the $Nonlinear$ layer, fliters $\bm{w_2}$ and bias $b_2$ in the $Conv2$ layer, and parameter $\tilde\eta$ in the $multi$ layer.

We compared our method to three state-of-the-art methods including SPIRiT, SAKE, and the recently released MoDL.  Quantitative metrics used included Peak Signal-to-Noise Ratio (PSNR), Structural Similarity (SSIM) and Normalized Mean Squared Error (NMSE).

\subsubsection{Results}
 We present a series of quantitative and qualitative reconstruction results, in which all the images were obtained by the direct square root of the multi-coil MR images reconstructed by different methods. 
\begin{figure}
	\centering
	\includegraphics[width=122mm, height=163mm]{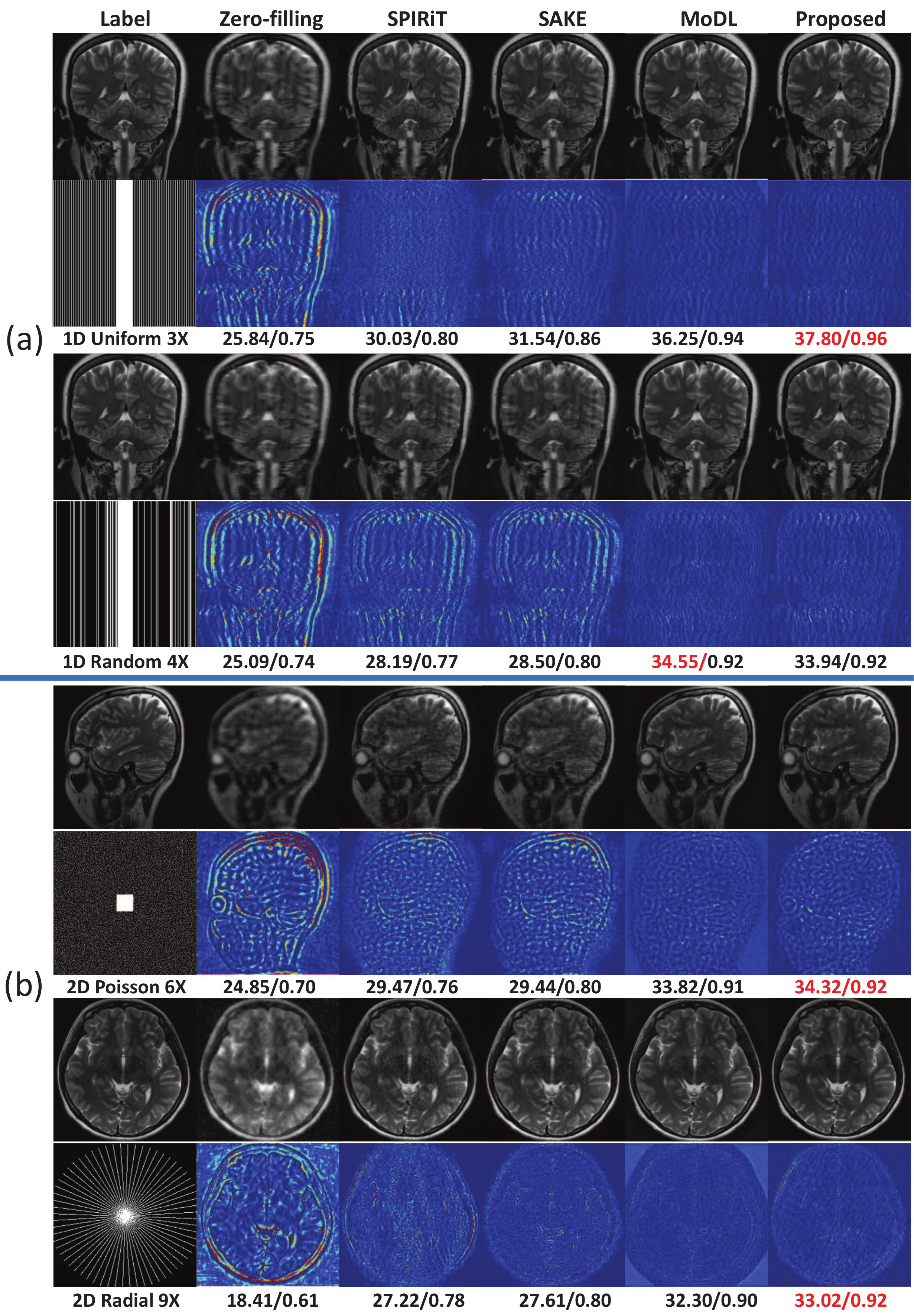}
	\caption{Comparison of different methods in reconstruction accuracy with different undersampling patterns and acceleration factors: reconstruction results and error maps are presented with corresponding quantitative measurements in PSNR/SSIM.} \label{fig3}
\end{figure}


Fig.~\ref{fig3} (a) shows the qualitative reconstruction results under a 1D uniform mask (ACS=28) with an acceleration factor of three, and a 1D random mask (ACS=24) with an acceleration factor of four. When the acceleration factor R=3, the SPIRiT method successfully removes structural artifacts, while the signal-to-noise ratio is low. On the contrary, the SAKE has a superior signal-to-noise ratio, but fails to remove the artifacts. The recently published MoDL method, like our method, has good reconstruction accuracy and robustness in removing artifacts and improving signal-to-noise ratio. Indeed, our method is slightly better than MoDL in terms of reconstruction error maps and metrics. Similar conclusions could be made for 1D random masks at the acceleration factor R=4. 

We also show the reconstruction results of 2D Poisson sampling with 6x acceleration and 2D radial sampling with 9x acceleration in Fig.~\ref{fig3} (b). It is apparent from the zero-filled images and the error maps that severe image detail loss occures under high acceleration conditions. Compared to the SPIRiT and SAKE methods, both our method and MoDL are capable of reconstructing high-quality images from highly undersampled k-space data. Interestingly, we found that the stability of MoDL is not as good as our method. In addition, from the error maps, the MoDL method has some noise at the edges of the image.
\begin{table}[tp]  
	
	\centering  
	\fontsize{6}{13}\selectfont  
	
	\caption{Quantitative metrics for pMRI reconstruction results (NMSE/PSNR/SSIM).}  
	\label{tab1}  
	\begin{tabular}{cc|c|c|c|c|c|c}  
		\hline  		
		&\bf Mask&\bf Rate&\bf Zero-{filling} &\bf SPIRiT&\bf SAKE\bf&\bf MoDL&\bf Proposed\bf\cr 
		\hline  
		&\ &3x&0.24/24.91/0.76&0.14/29.38/0.78
		&0.12/30.66/0.84
		&0.06/36.53/0.94
		&\bf0.06/36.99/0.96
		\cr  
		&1D Uniform&4x&0.28/23.58/0.72
		&0.16/28.25/0.74
		&0.17/27.93/0.80	
		&\bf0.07/35.60/0.93
		&0.09/33.56/0.93
		\cr
		&\ &5x&0.35/21.54/0.63
		&0.22/25.69/0.68		
		&0.25/24.59/0.71		
		&0.19/26.66/0.74		
		&\bf0.18/27.20/0.82
		\cr  			
		\hline
		&\ &3x&0.22/25.76/0.79
		&0.14/29.29/0.79		
		&0.13/29.99/0.84		
		&0.07/35.77/0.94		
		&\bf0.06/36.08/0.95		
		\cr  
		&1D Random&4x&0.26/24.28/0.74		
		&0.18/27.44/0.76		
		&0.18/27.58/0.80			
		&{\bf0.08/34.14}/0.92	
		&0.09/33.45/\bf0.92		
		\cr
		&\ &5x&0.30/22.96/0.69		
		&0.18/27.49/0.75				
		&0.19/26.92/0.78				
		&0.11/31.20/0.87				
		&\bf0.11/31.43/0.90		
		\cr   
		\hline
		&\ &4x&0.33/22.18/0.68
		&0.16/28.65/0.73		 
		&0.14/29.86/0.80		 
		&0.07/35.33/0.93		 
		&\bf0.07/35.54/0.94		 
		\cr  
		&2D Poisson&6x&0.35/21.55/0.65		 
		&0.20/26.43/0.66		 
		&0.19/27.15/0.74		 	
		&0.10/32.63/0.90		 
		&\bf0.10/32.64/0.90		 
		\cr
		&\ &9x&0.37/21.16/0.63		 
		&0.24/24.92/0.61		 		
		&0.23/25.00/0.68		 		
		&{\bf0.13/30.01}/0.85		 		
		&0.13/29.94/\bf0.86		 
		\cr   
		\hline
		&\ &4x&0.18/27.35/0.84
		&0.13/29.87/0.78		 
		&0.10/32.62/0.86		 
		&{0.06}/36.87/0.95	 
		&\bf0.06/37.02/0.96		 
		\cr  
		&2D Radial&6x&0.26/24.38/0.75		 
		&0.15/28.83/0.74		 
		&0.12/30.88/0.83		 
		&{\bf0.08/34.79}/0.93		 
		&0.08/34.72/\bf0.94		 
		\cr
		&\ &9x&0.37/21.29/0.64		 
		&0.19/27.02/0.72		 		
		&0.15/28.67/0.79		 		
		&0.11/31.62/0.88		 		
		&\bf0.10/32.49/0.91	 
		\cr   
		\hline
	\end{tabular}  
\end{table}

Table~\ref{tab1} summarizes the average quantitative results of 50 test images using  different methods, different undersampling patterns and different acceleration factors. We can see that using the same acceleration factor, 2D undersampling reconstruction results are better than 1D, which is expected. Our approach improves the average PSNR by nearly 5dB compared to the traditional SAKE and SPIRiT methods. Additionally, it can be observed that the quality of the reconstruction relies on the undersampling patterns. Under the same acceleration factor, the random mask is better than the uniform mask as the acceleration factor increases. Similarly, the radial mask is better than the Poisson mask.

\section{Conclusions}
In this paper, we propose a novel model-based convolutional de-aliasing network for fast pMRI. The proposed method explores the redundancy and correlation of parallel MR images with deep learning. A split Bregman iterative algorithm was developed to solve the proposed model. Experimental results demonstrated that our method achieved comparable and even superior reconstruction results than existing methods both quantitatively and qualitatively. In addition, the proposed method has another merit that does not require explicit estimation of coil sensitivity.

\subsubsection{Acknowledgement.}
This research was partly supported by the National Natural Science Foundation of China (61601450,61871371, 81830056), Science and Technology Planning Project of Guangdong Province (2017B020227012, 2018B01 0109009), the Basic Research Program of Shenzhen (JCYJ20180507182400762), Youth Innovation Promotion Association Program of Chinese Academy of Sciences (2019351).

\end{document}